\documentclass[12pt]{article}
\usepackage{amsmath}
\usepackage{amssymb}
\usepackage{bm}
\usepackage{graphicx}
\usepackage{enumerate}
\usepackage{natbib}
\usepackage{url} 
\usepackage{caption}
\usepackage{subcaption}
\usepackage{tikz}
\usetikzlibrary{calc}

\begin{document}

 \title{\bf Random graphical model of microbiome interactions in related environments}
  \author{Veronica Vinciotti\\
    Department of Mathematics, University of Trento,\\
Ernst C. Wit and Francisco Richter \\
    Institute of Computing, Universit\`a della Svizzera italiana
		}
		\date{}
\maketitle
\begin{abstract}{The microbiome constitutes a complex microbial ecology of interacting components that regulates important pathways in the host. Measurements of microbial abundances are key to learning the intricate network of interactions amongst microbes. Microbial communities at various body sites tend to share some overall common structure, while also showing diversity related to the needs of the local environment.  We propose a computational approach for the joint inference of microbiota systems from metagenomic data for a number of body sites. The random graphical model (RGM) allows for heterogeneity across the different environments while quantifying their relatedness at the structural level. In addition, the model allows for the inclusion of external covariates at both the microbial and interaction levels, further adapting to  the richness and complexity of microbiome data. Our results show how: the RGM approach is able to capture varying levels of structural similarity across the different body sites and how this is supported by their taxonomical classification; the Bayesian implementation of the RGM fully quantifies parameter uncertainty; the microbiome network posteriors show not only a stable core, but also interesting individual differences between the various body sites, as well as interpretable relationships between various classes of microbes.}
\end{abstract}

\section{Introduction}
The microbiome constitutes a complex microbial ecology of interacting components that regulates important pathways in the host. Microbiotic systems have been intensively studied in recent years, and associations have been found with a number of health conditions, such as obesity \citep{lechatelier13}, diabetes \citep{pedersen16} and the response to immunotherapy \citep{lee22}. Rich sources of high-throughput data of the microbiome, such as those generated by the Human Microbiome Project \citep{HMP} and the Metagenomics of the Human Intestinal Tract (MetaHIT) project  \citep{qin10} are key to learning the intricate network of interactions amongst microbe communities. 

As the microbiome interacts with the local environment, the microbiome varies in constitution profile at different body sites \citep{sharon22}. 
For example, \cite{segata12} find four groups of digestive tract sites, characterized by distinct bacterial compositions and metabolic processes. 
Despite this heterogeneity, from a structural perspective, it is expected that the interaction profile is largely shared between different body sites. This constitutes a core microbiome network, describing stable components of the microbiome interactions across time, body sites and populations.

Most studies rely solely on fecal samples to represent the gut microbiome and on  saliva samples to describe the oral microbiome \citep{sharon22}. Equally, all available methods and implementations, such as the commonly used SparCC \citep{friedman12} and SPIEC-EASI \citep{kurtz15}, infer one microbiota system from a given dataset. As such, they are suited to learn either environment-specific systems from a study on that environment or a core microbiome network from pooled data across different body sites.  Instead, we propose a computational approach for the joint inference of microbiota systems from metagenomic data for a number of body sites that captures both the core metabolic network as well as individual differences. 

Closest to the approach proposed in this paper, \cite{vinciotti22} develop a Gaussian copula graphical model to infer microbiota systems from count genomic data. While the approach recovers a core microbiome network generating the data, the parametric form used for the marginals is able to capture both the heterogeneity of microbial abundances across different body sites and the typical features of microbial data, such as zero inflation and compositionality. An efficient Bayesian inferential procedure allows to quantify the uncertainty around the estimated parameters. This is the case also for the recovered network, i.e., interactions between microbes are returned with an associated posterior edge probability and partial correlation distribution. 

In this paper, we aim to capture heterogeneity also at the structural level of microbial interactions, while quantifying the possible relatedness among microbiota systems from different environments. To this end, we augment the model of \cite{vinciotti22} with a random graph model on the conditional independence graphs that describe the joint microbial count distributions at each body site. We define a novel  \textit{random graphical model} as the combination of a graphical model with this random independence graph model.  

Borrowing from the network science literature \citep{hoff02}, we formalise the random graph model as a latent probit network model, where the probability of an edge in a particular microbiota system depends on a latent space of potentially related environments, i.e., it will increase if the body site is close in this latent space to another environment where that particular edge is present. In addition, the edge probability depends on individual network sparsity levels for each body site and on external covariates at the network level. For the latter, we consider the effect of taxonomy sharing on the propensity of microbes to interact, but, in principle, any other coavariate or external knowledge can be included at this stage.  

We apply the proposed random graphical model to a microbiome study, which is part of the Human Microbiome Project (HMP). After preprocessing, we use 4500 samples of 87 microbes across 13 body sites. Our analysis shows that the latent space is able to capture the biological relatedness between the 13 microbiotic systems. Indeed, the locations of the body sites in the inferred latent space match closely both with the classification made by \cite{segata12} and with the Uberon anatomy classification of body sites \citep{mungall12}. The environment-specific networks, and in particular their associated estimated edge probabilities, can be queried further, in order to characterize the individual networks as well as to highlight commonalities and differences between the 13 environments. Beyond the information that can be discovered from the data using the proposed model, we find that the new approach leads to a more stable recovery of the microbiotic systems, compared to individual analyses conducted for each body site separately. 

In Section~\ref{sec:methods}, we describe the random graphical model and the Bayesian inference procedure in detail, while in Section~\ref{sec:results} we present the results on real and simulated data, before a discussion and conclusion in Section~\ref{sec:conclusion}.

\section{Methods}\label{sec:methods}
\subsection{Random graphical model}
In this section we define the random graphical model for network inference from heterogeneous microbiome data from a number of environments. Let $\mathbf{Y}^{(k)}=(Y^{(k)}_1,\ldots,Y^{(k)}_p)$ be the random vector of interest, that is the abundances of $p$ Operating Taxonomic Units (OTUs) in environment $k$, with $k=1, \ldots,B$. In our study, $B=13$. At an abstract level, we assume $\mathbf{Y}^{(k)}$ to be distributed according to some graphical model (GM),
\[ \mathbf{Y}^{(k)}|G^{(k)}\sim \mathcal{L}_{G^{(k)}}(\boldsymbol{\Omega}^{(k)}),\]
relative to some conditional independence graph $G^{(k)}$ with some associated parameters $\boldsymbol{\Omega}^{(k)}$. The graphs $G=\left\{G^{(k)}\right\}_k$ are themselves  distributed according to a joint random graph model, 
\[ G^{(k)}\sim P(\boldsymbol{\Theta}),~~ k=1,\ldots,B\]
for some vector of parameters $\boldsymbol{\Theta}$. 

The type of graphical model and the type of random graph model depend on the situation under consideration. As for the graphical model, we consider the Gaussian copula graphical model, due to its easy mathematical formulation. It is a flexible model for multivariate non-Gaussian data, as the count microbiome data under consideration. Thus, similarly to \citep{cougoul19,vinciotti22}, we assume:
\[P(Y^{(k)}_{1} \leq y_{1},\ldots, Y^{(k)}_{p} \leq y_{p}~|~G^{(k)}, \boldsymbol{\Omega}^{(k)}) = 
\Phi_p \big( \Phi^{-1}(F_1(y_1)), \ldots, \Phi^{-1}(F_p(y_p); \mathbf{R}^{(k)})  \big),
\]
where $\Phi_{p}$ is the cumulative distribution function of a $p$-dimensional multivariate normal with a zero mean vector and  correlation matrix $\mathbf{R}^{(k)} = {\boldsymbol{\Omega}^{(k)}}^{-1}$, $\Phi$ is the standard univariate normal distribution function, and $F_j$ is the marginal distribution of OTU $j$. The dependency structure induced by this model in condition $k$ is represented by the conditional independence graph $G^{(k)}$. Following from the theory of Gaussian graphical models \citep{lauritzen96}, this is given by the zero-patterns of the inverse of the correlation matrix $\boldsymbol{\Omega}^{(k)}$, typically called the precision or concentration matrix. 

In order to adapt to the richness and heterogeneity of microbiome data, the marginal distribution of OTU $j$ can be linked to external covariates, including body site and normalizing factors such as sequencing depth. As in \citep{vinciotti22}, we formalise this with the use of a parametric marginal model. In particular, we consider discrete Weibull regression marginals \citep{peluso19}, i.e., $j=1,\ldots,p$,
\begin{align}
\label{eq:dwmarginals}
F_{j}(y_j|\bm{x}) &= 1-q_j(\bm{x})^{(y_j+1)^{b_j(\bm{x})}}\\ \nonumber
\log\left(\frac{q_j(\bm{x})}{1-q_j(\bm{x})}\right) &= \mathbf{x}^t \boldsymbol{\eta}_j, \quad \log\left(b_j(\bm{x}) \right) = \bm{x}^t \boldsymbol{\gamma}_j
\end{align}
with node covariates $\bm{x}=(1,x_1,\ldots,x_m)^\top$ and regression coefficients $\boldsymbol{\eta}_j$ and $\boldsymbol{\gamma}_j$ associated to the two parameters defining the discrete Weibull distribution, respectively. Due to the discreteness of the data, the mapping to the latent Gaussian space $z_j=\Phi^{-1}(F_j(y_j))$ of the copula is not unique. Indeed, each observation $(y_j,\bm{x})$ is associated to an interval in the latent space, given by
\begin{equation}
\label{eq:dwintervals}
\mathcal{I}_{F_j}(y_j|\bm{x})  = \big(\Phi^{-1} \big(F_j(y_{j}-1|\bm{x}) \big),\Phi^{-1} \big(F_j(y_{j}|\bm{x}) \big)\big].
\end{equation}

As for the joint random graph model, we are particularly interested in modelling the relatedness of the different environments as well as a possible link with external covariates/existing knowledge at the microbial interaction level. To this end, we formalise the model with the following latent probit network model \citep{hoff02}
\begin{equation}
\label{eq:latentprobit}
P({G_{j_1,j_2}}^{(k)}=1~|~G_{j_1,j_2}^{(-k)}, \boldsymbol{\Theta}, \bm{w})= 
 \Phi\Big(\alpha_k+{\bm{w}_{j_1,j_2}}^t\boldsymbol\beta+\mathbf{c}_k^t\sum_{k' \ne k}\mathbf{c}_{k'}1_{\{{G_{j_1,j_2}}^{(k')}=1\}}\Big),
\end{equation}
where 
${G_{j_1,j_2}}^{(k)}=1$ defines an edge between node $Y_{j_1}$ and node $Y_{j_2}$ in condition $k$, $\bm{w} \in \mathbb{R}^d$ is the vector of edge-specific covariates,  $\mathbf{c}_1,\ldots,\mathbf{c}_B \in \mathbb{R}^2$ are the latent space variables for each condition, $\alpha_k$ is the intercept of the model and relates to the overall sparsity level of graph  $G^{(k)}$. We denote with $\boldsymbol{\Theta}=(\boldsymbol{\alpha},\boldsymbol{\beta},\mathbf{c})$ the vector of parameters associated to the joint random graph model.

In the next section, we discuss inference of the full set of model parameters from microbiome data, namely $\boldsymbol{\eta}_1, \ldots, \boldsymbol{\eta}_p, \boldsymbol{\gamma}_1, \ldots, \boldsymbol{\gamma}_p$ at the marginal level and $G^{(1)},\ldots, G^{(B)}$, $\boldsymbol{\Omega}^{(1)}, \ldots, \boldsymbol{\Omega}^{(B)}, \boldsymbol{\Theta}$ at the structural level.

\subsection{Bayesian inference} \label{sec:bayesalg}
\begin{figure}[!tpb]
\begin{center}
\begin{tikzpicture}
\node (a) at (-1,5.5) {$\boldsymbol{\Theta}=(\boldsymbol{\alpha},\boldsymbol{\beta},\mathbf{c}), \: \bm{w}$};
\node (b)  at (-1,3) {$G^{(1)},\ldots,G^{(B)}$};
\node (c) at (-1,1.5) {$\boldsymbol{\Omega}^{(1)},\ldots, \boldsymbol{\Omega}^{(B)}$};
\node (d) at (-1,-0.5) [text width=3cm] {${\bm{z}_1}^{(k)},\ldots,{\mathbf{z}_n}^{(k)}$ \\ $k=1,\ldots,B$};
\node (e) at (3,-0.5) [text width=4cm] {$\boldsymbol{\eta}_j,\boldsymbol{\gamma}_j$, $j=1,\ldots,p$ (${F_1}^{(k)},\ldots,{F_p}^{(k)}$)};
\node (f) at (1,-2.5) [text width=2.5cm] {${\mathbf{y}_1}^{(k)},\ldots,{\bm{y}_n}^{(k)}$ $k=1,\ldots,B$};
\draw (a) edge[->,line width=2pt] node[anchor=center, above right, text width=8cm] {$G^{(k)} \sim \mbox{Bernoulli}(P(G^{(k)}=1~|~G^{(-k)}, \boldsymbol{\Theta}, \bm{w}))$} (b);  
\draw (b) edge[->,line width=2pt] node[anchor=center,  right, text width=5cm] {$\boldsymbol{\Omega}^{(k)} \sim W_G(3,\mathbb{I}_p)$} (c);
\draw (c) edge[->,line width=2pt] node[anchor=center,  right, text width=5cm] {$\mathbf{{z}_i}^{(k)} \sim N_p(\mathbf{0},{\boldsymbol{\Omega}^{(k)}}^{-1})$} (d);
\draw (d) edge[->,line width=2pt] (f);
\draw (e) edge[->,line width=2pt] (f);
\draw[black,thick] ($(a)+(-2,0.3)$) rectangle ($(b)+(10,0.9)$);
\node (g) at ($(a)+(6.3,0)$) {\textbf{latent network probit model}};
\draw[black,thick] ($(b)+(-2,0.6)$) rectangle ($(f)+(8,-0.6)$);
\node (g) at ($(b)+(6.5,0.3)$) {\textbf{Gaussian copula GM}};
\end{tikzpicture}
\end{center}
\caption{Sketch of the random graphical model generating microbiome data across different related environments. The model combines a latent network probit model with a Gaussian copula graphical model.}\label{fig:01}
\end{figure}
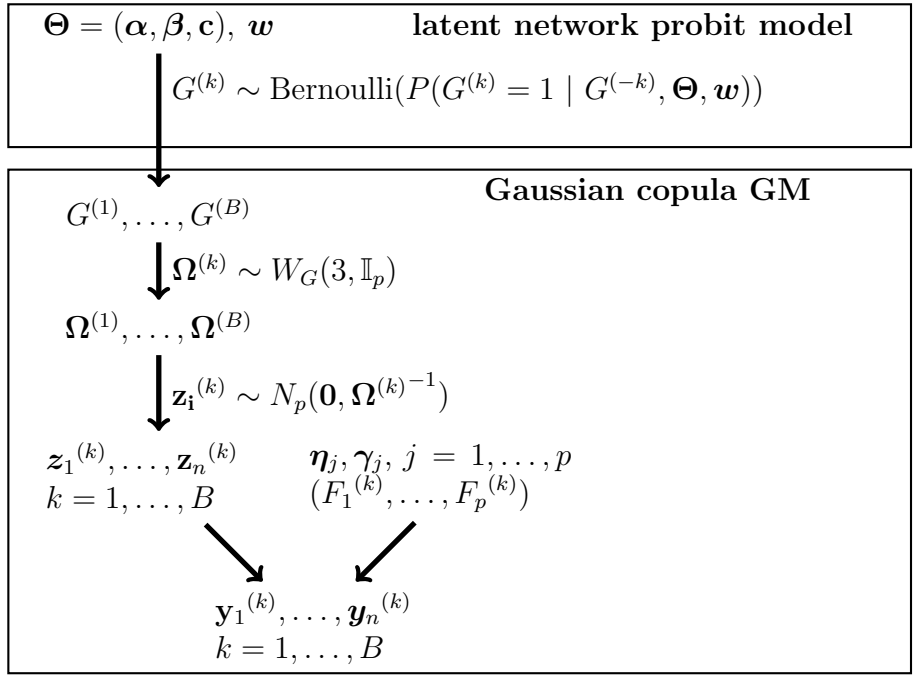
Figure \ref{fig:01} describes the proposed random graphical model and how it generates microbiome data from different, possibly related, environments. In particular, given the parameters $\boldsymbol{\Theta}=(\boldsymbol{\alpha},\boldsymbol{\beta},\mathbf{c})$ that define the latent network model, edges in $G^{(k)}$ become independent, conditional on the remaining graphs, and are, therefore, the result of Bernoulli draws. Given graphs $G^{(k)}$ for each condition $k=1,\ldots,B$, the data are then generated via a Gaussian copula graphical model \citep{vinciotti22}, i.e., positive-definite precision matrices are drawn via G-Wishart distributions, and the resulting multivariate normal draws are combined with the parametric marginals to generate count microbiome data.  

In order to quantify the full uncertainty in the estimation of the parameters, we opt for a Bayesian inferential procedure. To this end, we consider the following prior distributions: $N(0,10)$ priors on each parameter in $\boldsymbol{\Theta}$, 
G-Wishart priors for the precision matrix $\boldsymbol{\Omega}^{(k)} \sim W_G(3,\mathbb{I}_p)$ conditional on the graph $G^{(k)}$, and $N(0,1)$ priors on each regression coefficient in ($\boldsymbol{\eta}_j,\boldsymbol{\gamma}_j$), $j=1,\ldots,p$. 

We have devised a Markov Chain Monte Carlo (MCMC) scheme for generating samples from the posterior distribution of the parameters. In particular, the scheme is made of the following steps:
\begin{enumerate}
\itemsep1.5em
\item Metropolis-Hastings sampling of the marginal regression parameters $\boldsymbol{\eta}_j,\boldsymbol{\gamma}_j$ \citep{haselimashhadi18} 
\item Gibbs sampling of the parameters $\boldsymbol{\Theta}$ from their posterior distribution via a sequence of probit regressions (with offset):
\begin{itemize}
\item $\boldsymbol{\alpha} | \boldsymbol{\beta},\mathbf{c}, \{G^{(k)}\}_k, \bm{w}$
\item $\boldsymbol{\beta} | \boldsymbol{\alpha},\mathbf{c}, \{G^{(k)}\}_k, \bm{w}$
\item $\mathbf{c}_k | \boldsymbol{\alpha},\boldsymbol{\beta},\mathbf{c}_{-k}, \{G^{(k)}\}_k, \bm{w}$
\end{itemize}
\item Gibbs sampling of ${z_{ij}}^{(k)} | \boldsymbol{\Omega}^{(k)}, {\bm{z}_{i,-j}}^{(k)}, {\bm{y}_i}^{(k)}$ via a truncated normal on $ \mathcal{I}_{\hat F_j}(y_{ij}| \bm{x}_i)$ \citep{vinciotti22}
\item Gibbs sampling of $\boldsymbol{\Omega}^{(k)} | G^{(k)},\bm{z}^{(k)}$ via a G-Wishart posterior distribution \citep{mohammadi15}
\item Continuous time birth-death MCMC sampling of the graph
\[{(G^{(k)})}^{\pm e}| \boldsymbol{\Omega}^{(k)}, \bm{z}^{(k)}, G^{(k)},\boldsymbol{\Theta}, \bm{w}\]
generating a new graph from the current graph $G^{(k)}$ with an edge $e$ added, removed or kept \citep{mohammadi15}.  
\end{enumerate}

Upon convergence, posterior distributions of all parameters are returned. We focus particularly on the parameters $\boldsymbol{\Theta}$, which provide information on the latent process generating the graphs and how related the different environments are at the structural level, and on the graphs $G^{(k)}$, which are associated to posterior edge inclusion probabilities
\begin{equation}
P({G_{j_1j_2}}^{(k)}=1|\bm{y},\bm{x},\bm{w}) = \frac{\sum_{t=1}^{N} 1((j_1,j_2) \in {G_t}^{(k)}) W({\bm{\Omega}_t}^{(k)},\boldsymbol{\Theta})}{\sum_{t=1}^{N} W({\bm{\Omega}_t}^{(k)},\boldsymbol{\Theta})},
\label{eq:posteriorprob}
\end{equation}
where $N$ is the number of MCMC iterations and $W({\bm{\Omega}_t}^{(k)},\boldsymbol{\Theta})$ is the waiting time for graph ${G_t}^{(k)}$ with precision matrix ${\bm{\Omega}_t}^{(k)}$ \citep{vinciotti22}. Posterior distributions on the precision matrices ${\bm{\Omega}}^{(k)}$ are also available and can be converted to partial correlations for each edge, via
\begin{equation}
\pi_{j_1j_2} = -\frac{\omega_{j_1j_2}}{\sqrt{\omega_{j_1j_1}\omega_{j_2j_2}}}.
\label{eq:parcor}
\end{equation}
These values give information also about the sign of the dependencies in each environment. In a similar vein, posterior distributions of any network statistic of interest can be derived from the MCMC chain of graphs that is returned.

\section{Results} \label{sec:results}
\subsection{Simulation study}
In order to clarify the data generating process behind the proposed random graphical model (Figure \ref{fig:01}) and to assess its performance in inferring parameters from data, we simulate $n=346$ observations on $p=87$ variables for $B=13$ environments, with the sample size and dimensions matching those of the real data. For the simulation, we construct a latent space $\boldsymbol{\Theta}$ with the following components: $\bm{\alpha}$ parameters drawn from a $N(-2,1)$ distribution, i.e., a high level of network sparsity; one edge covariate $W$ from a $U(-0.5,0.5)$ distribution with an associated parameter $\beta=2.5$; latent vectors $\mathbf{c}\in \mathbb{R}^2$ with each component drawn from a $N(0,0.3)$ distribution. Given $\boldsymbol{\Theta}$, we first generate 13 graphs $G^{(k)}$ via Bernoulli draws for each edge conditional on the others. We repeat the sampling for 100 iterations in order to sample from the actual joint distribution of the graphs. Given the true graphs, we sample their associated precision matrices from a $W_G(3,\mathbb{I}_p)$ distribution and we finally obtain the data for each environment from a multivariate Gaussian draw. We omit here the case of discrete marginals and concentrate on the inference of the latent space and recovery of the networks. The data constructed as described above can be retrieved by running the function \texttt{sim.rgm} of the \texttt{rgm} package accompanying this paper, using the default values for the inputs.

\begin{figure*}[!tpb]
\centering
\includegraphics[scale=0.45]{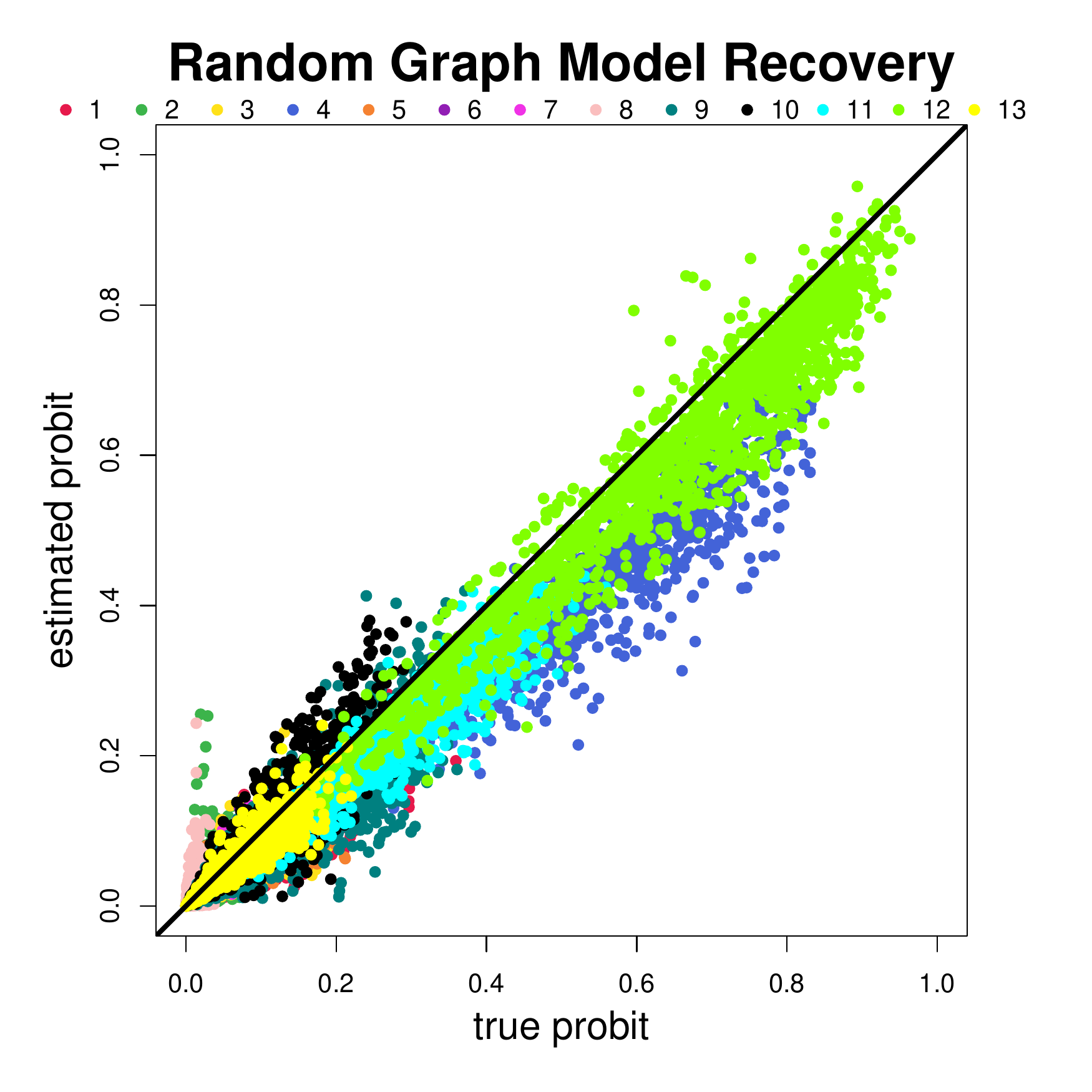}
\includegraphics[scale=0.45]{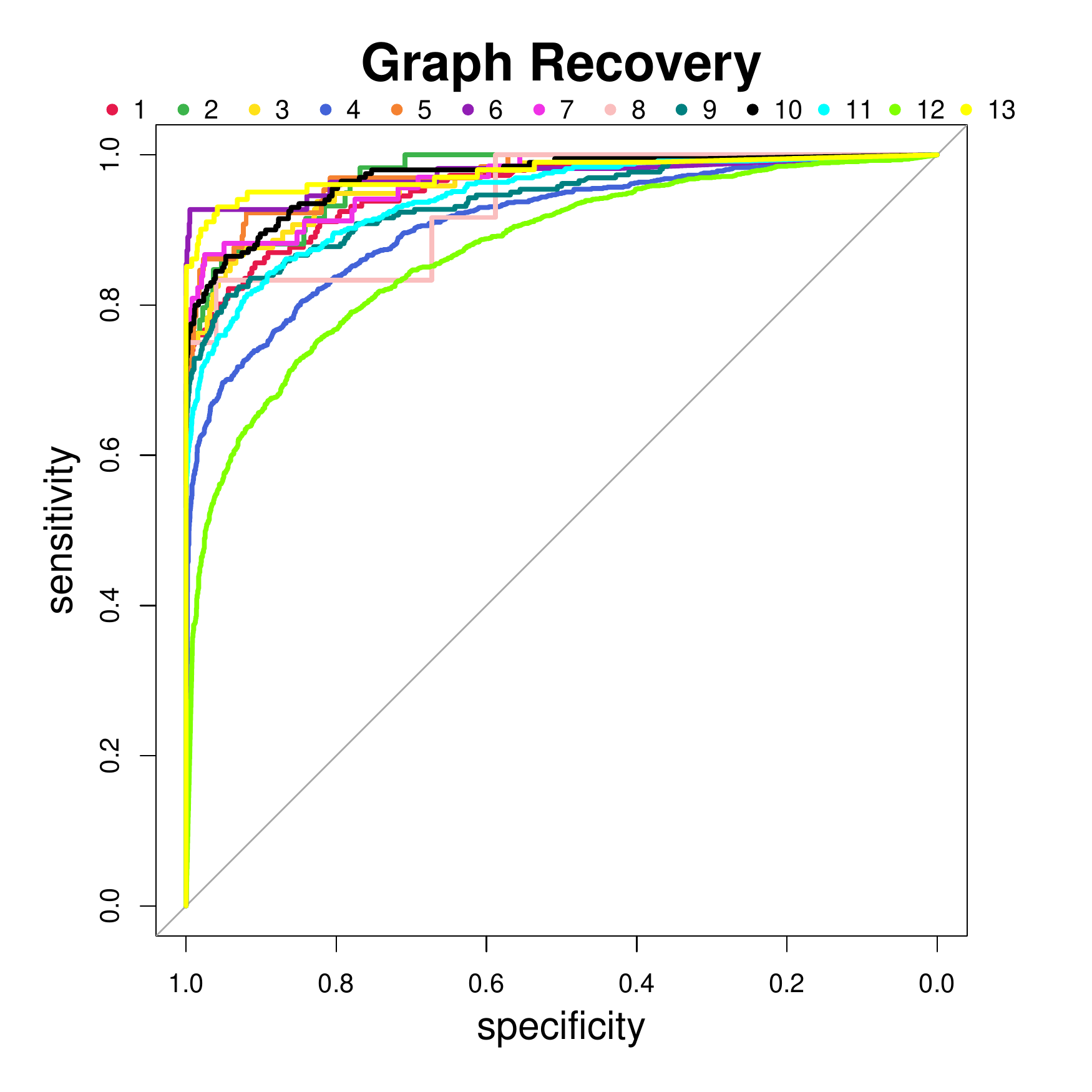}
\caption{Results from the simulation study, evaluated on the last 2500 MCMC iterations. Left:  True probit probabilities from Equation (\ref{eq:latentprobit}) versus those calculated using the mean posterior estimates of the $\boldsymbol{\alpha}$, $\beta$ and $\mathbf{c}$ parameters. Right: Receiver operating characteristic curves of the recovered graphs against the true ones for each environment, across a sequence of thresholds on the posterior edge probabilities from Equation (\ref{eq:posteriorprob}). The 13 colours distinguish the 13 environments, respectively. \label{fig:02}}
\end{figure*}

Figure \ref{fig:02} reports the results after 10000 MCMC iterations, obtained by running the function \texttt{rgm} with prior distributions as described in Section \ref{sec:bayesalg}. The computational time is about 14 seconds per iteration. We retain the last 25\% of the iterations for the calculation of posterior edge distributions from Equation (\ref{eq:posteriorprob}) and posterior distributions of the  parameters $\boldsymbol{\Theta}$ of the random graph model. The first plot shows a good recovery of the latent network space $\boldsymbol{\Theta}$, by comparing the true probit probabilities from Equation (\ref{eq:latentprobit}) with those obtained using the mean posterior estimates of the $\boldsymbol{\alpha}$, $\beta$ and $\mathbf{c}$ parameters. The second plot shows an accurate reconstruction of the networks ${G}^{(k)}$, by comparing the recovered graphs with the true graphs, for each environment and across a sequence of thresholds on the posterior edge probabilities. The average area under the receiver operating characteristic curves is 0.95,  across the 13 environments.

\subsection{Joint inference of microbiota systems across body sites}

We use the microbiome data from the \texttt{rMAGMA} package \citep{cougoul19}, collecting microbial abundances at the level of Operating Taxonomic Units (OTUs) from 16S variable region V3-5 data of healthy individuals from the Human Microbiome Project \citep{HMP}. After filtering out samples with less than 500 reads, we focus on the 13 body sites with the largest sample size, namely  ``Anterior\_nares'' (later referred to as \texttt{nose}),   ``Attached\_Keratinized\_gingiva'' (\texttt{ker-gingiva}), ``Buccal\_mucosa'' (\texttt{cheek}), ``Hard\_palate'' (\texttt{palate}), ``L\_Retroauricular\_crease'' (\texttt{L-ear}), ``Palatine\_Tonsils'' (\texttt{tonsils}), ``R\_Retroauricular\_crease'' (\texttt{R-ear}), ``Saliva'' (\texttt{saliva}), ``Stool'' (\texttt{stool}), ``Subgingival\_plaque''  (\texttt{sub-gingiva}), ``Tongue\_dorsum'' (\texttt{tongue}), ``Throat'' (\texttt{throat}), ``Supragingival\_plaque'' (\texttt{sup-gingiva}). On average, there are $346$ samples for each body site. We finally restrict our attention to the 87 OTUs which have more than two distinct observed values in each of these environments. The microbal communities are the interacting units, and, therefore, constitute the nodes of the network. 

A number of covariates are considered both at the marginal level of each OTU and at the network level. As covariates $\bm{x}$ for the discrete Weibull marginal distribution for each OTU (Equation \ref{eq:dwmarginals}), we include the library size for each sample, estimated by the geometric mean of
pairwise ratios of OTU abundances of that sample with all other samples (function \texttt{GMPR} in \texttt{rMAGMA}), dummy variables for each body site, and interactions between body sites and library size. This results in 26 parameters per OTU. As regards to the random graph model in Equation \ref{eq:latentprobit}, this is defined by a sparsity parameter $\alpha_k$ and a latent location $\bm{c}_k \in \mathbb{R}^2$, for each environment, and a vector $\boldsymbol{\beta}$ of regression coefficients associated to six binary variables ($\bm{w}$) that encode the belonging of a pair of OTUs to the same taxonomy level. In particular, we consider the six taxonomy levels given, respectively, by the bacterial \texttt{phylum}, \texttt{class}, \texttt{order}, \texttt{family}, \texttt{genus} and \texttt{species}.

\begin{figure}[!tpb]
\centering
\includegraphics[scale=0.45]{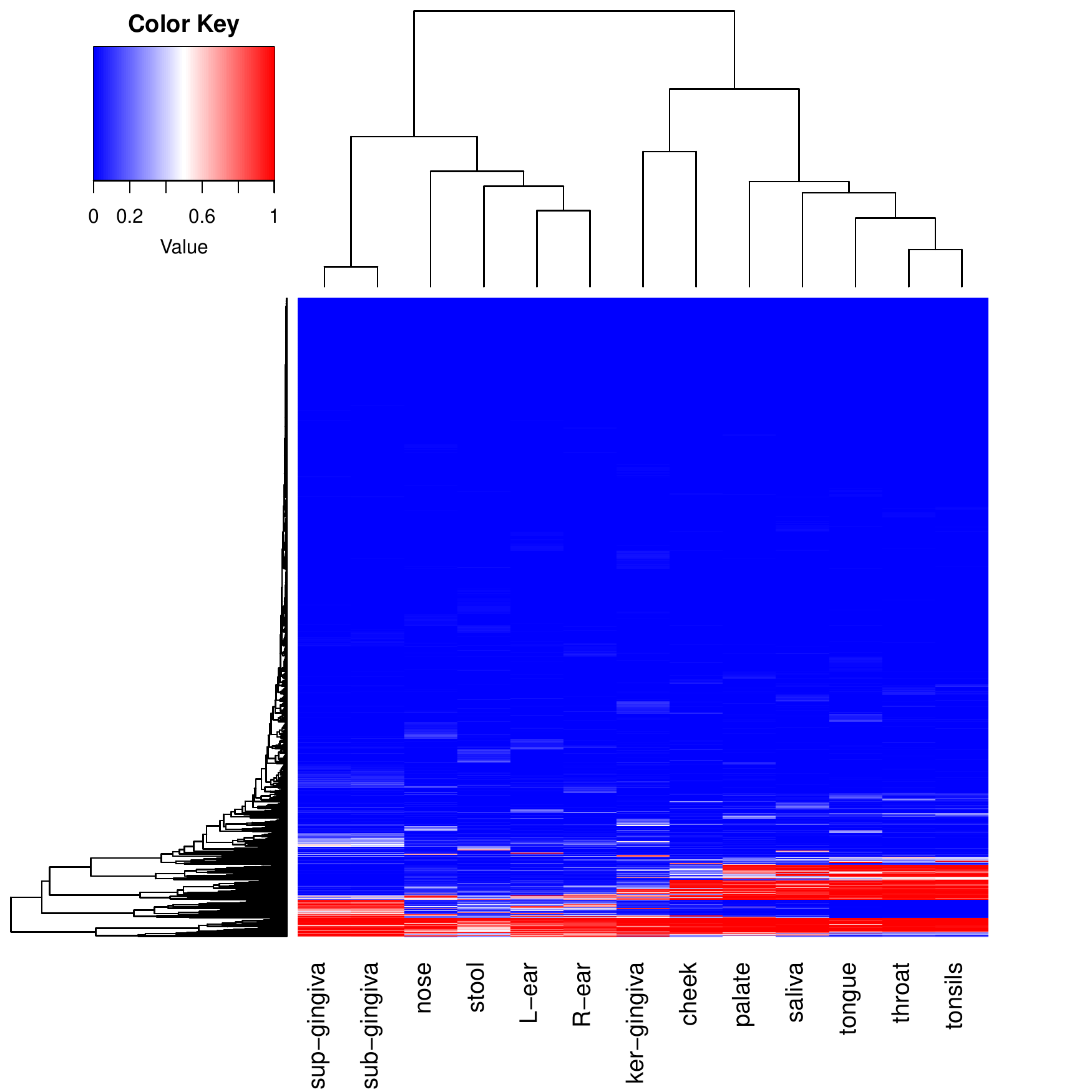}
\caption{Random graphical model inferred from microbiome data: Posterior edge probabilities for each edge and each environment, rearranged via row and column clustering. \label{fig:03}}
\end{figure}

We fit discrete Weibull parametric marginals for each OTU  via 50000 MCMC iterations (function \texttt{bdw.reg} in the \texttt{BDgraph} package \citep{mohammadi19}). As typical of inferential approaches for Gaussian copula graphical models, this step is performed first, followed by a calculation of the intervals from Equation \ref{eq:dwintervals} using posterior mean estimates of the parameters (evaluated on the last 25\% of the iterations). These intervals are then used for the subsequent learning of the structural dependencies, by iterating steps 2-5 of the procedure described in Section \ref{sec:bayesalg} (function \texttt{rgm} in the \texttt{rgm} package that accompanies this paper). 
Given the huge space of graphs, we let the Bayesian structural learning procedure run for 3 million MCMC iterations. All subsequent results are evaluated on the last 7,500 iterations. 
\vskip .1in
\noindent
{\bf Interpretation of results.}
The most immediate output of the analysis are the 13 networks that are inferred for each environment. Figure \ref{fig:03} summarises these networks by the posterior edge probabilities, calculated from Equation (\ref{eq:posteriorprob}). It is clear that the networks tend to be sparse, and vary to some extend between the conditions.  
The reasons for this environmental network variation can be found from the random graph generative process, described by Equation (\ref{eq:latentprobit}). Figure \ref{fig:03} shows a high level of sparsity across all networks (mean posterior edge probabilities equal to 6.8\%, on average across the 13 environments). This is captured by low $\alpha$ values of the fitted random graph model (mean posterior estimate -2.7, on average across the 13 environments). 

\begin{figure}[!tpb]
\centering
\includegraphics[scale=0.45]{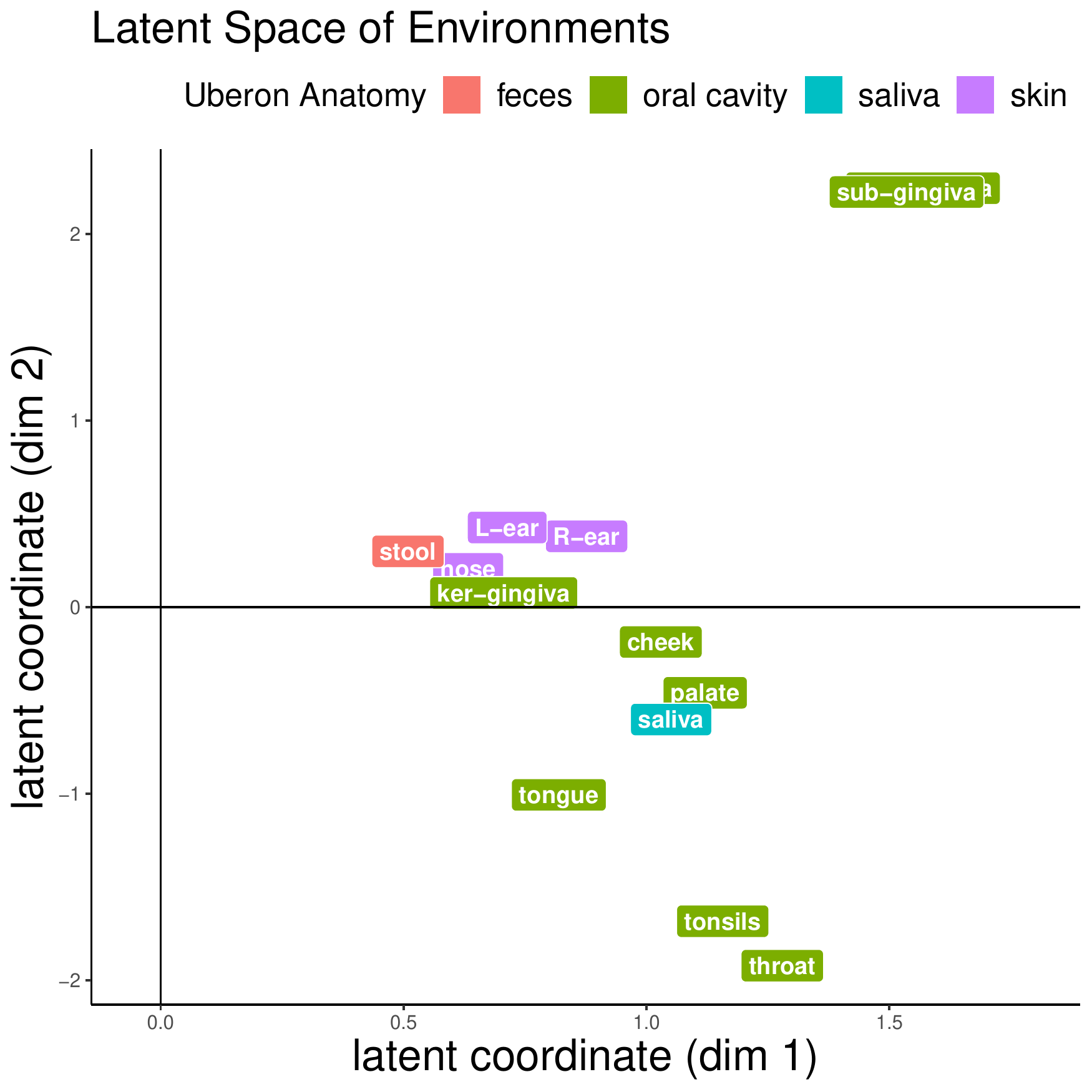}
\caption{Random graphical model inferred from microbiome data: mean posterior locations of the body sites ($\bm{c}$) in a 2D latent space. Colours refer to the Uberon anatomy classification of the body sites.
\label{fig:04}}
\end{figure}

Figure \ref{fig:03} shows a high structural similarity between some environments, with a high sharing of edges with high probability and of edges with low probability among similar environments. This is captured by the latent locations $\bm{c}$ of the random graph model, whose posterior means are plotted in Figure \ref{fig:04}. For example, \texttt{sup-gingiva} and \texttt{sub-gingiva} are highly related environments, and similarly \texttt{throat} and \texttt{tonsils}. Indeed, in both cases the two associated $\bm{c}$ vectors have a large inner product,  as they are close to each other in the space and far from zero. The indicator function in Equation (\ref{eq:latentprobit}) further encourages sharing of edges between these networks. Indeed, 93\% of the edges with posterior edge probability greater than 0.5 are in common between the \texttt{sup-gingiva} and \texttt{sub-gingiva} networks, and 95\% between the \texttt{throat} and \texttt{tonsils} networks. Looking at the posterior mean of partial correlations, calculated from the precision matrices via Equation (\ref{eq:parcor}), we find an agreement also on the sign of the dependency, with a correlation of 0.90 between \texttt{sup-gingiva} and \texttt{sub-gingiva} partial correlation values for each edge, and 0.93  between \texttt{throat} and \texttt{tonsils}. Finally, as the two pairs of networks are almost orthogonal to each other in the latent space, we expect little structural sharing between the \texttt{sup-gingiva}/\texttt{sub-gingiva} networks and the \texttt{throat}/\texttt{tonsils} networks. Indeed, they have the lowest agreement of high probability edges across all pairs, with an average sharing of 21.3\%. 

The similarities between the environments detected by the proposed method are partly supported by the Uberon anatomy classification of body sites, particularly when it comes to the three skin-related body sites. These are located close to each other in the latent space of Figure \ref{fig:04} and have on average 63\% of high-probability edges in common (Figure \ref{fig:03}). On the other hand, the oral cavity-related body sites appear to be further split into two groups. This is in line with the analysis of \cite{segata12}, which find four groups of body sites based on similar community compositions, namely: \texttt{cheek}, \texttt{ker-gingiva}, \texttt{palate}; \texttt{saliva}, \texttt{tongue}, \texttt{tonsils}, \texttt{throat}; \texttt{sub-gingiva} and \texttt{sup-gingiva}; and
\texttt{stool}. These groups are also clearly evident in Figure~\ref{fig:04}.

\begin{figure}[!tpb]
\centering
\includegraphics[scale=0.45]{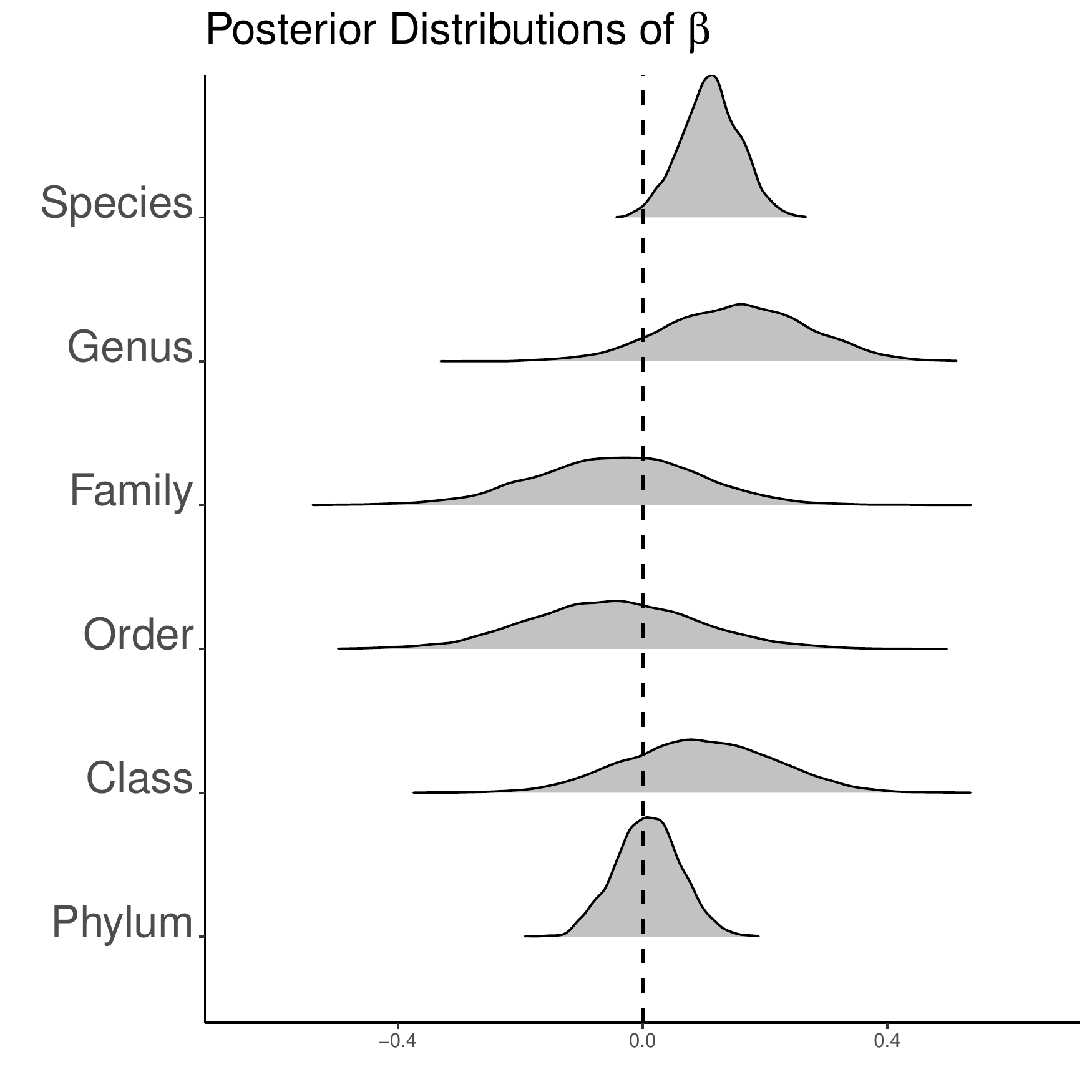} 
\caption{Random graphical model inferred from microbiome data: estimation of $\boldsymbol{\beta}$ parameters associated to six dummy variables indicating if a pair of nodes forming an edge belongs to the same taxonomy level 
\label{fig:05}}
\end{figure}

Finally, the taxonomical relatedness of the microbes encourages the presence of a link between them. Figure \ref{fig:05} shows how the probability of two OTUs connecting, in any environment, is positively associated with their belonging to some of the taxonomy levels considered, in particular to the \texttt{species}, \texttt{genus} and \texttt{class} taxonomies.

\begin{figure}[!tpb]
\centering
\includegraphics[scale=0.4]{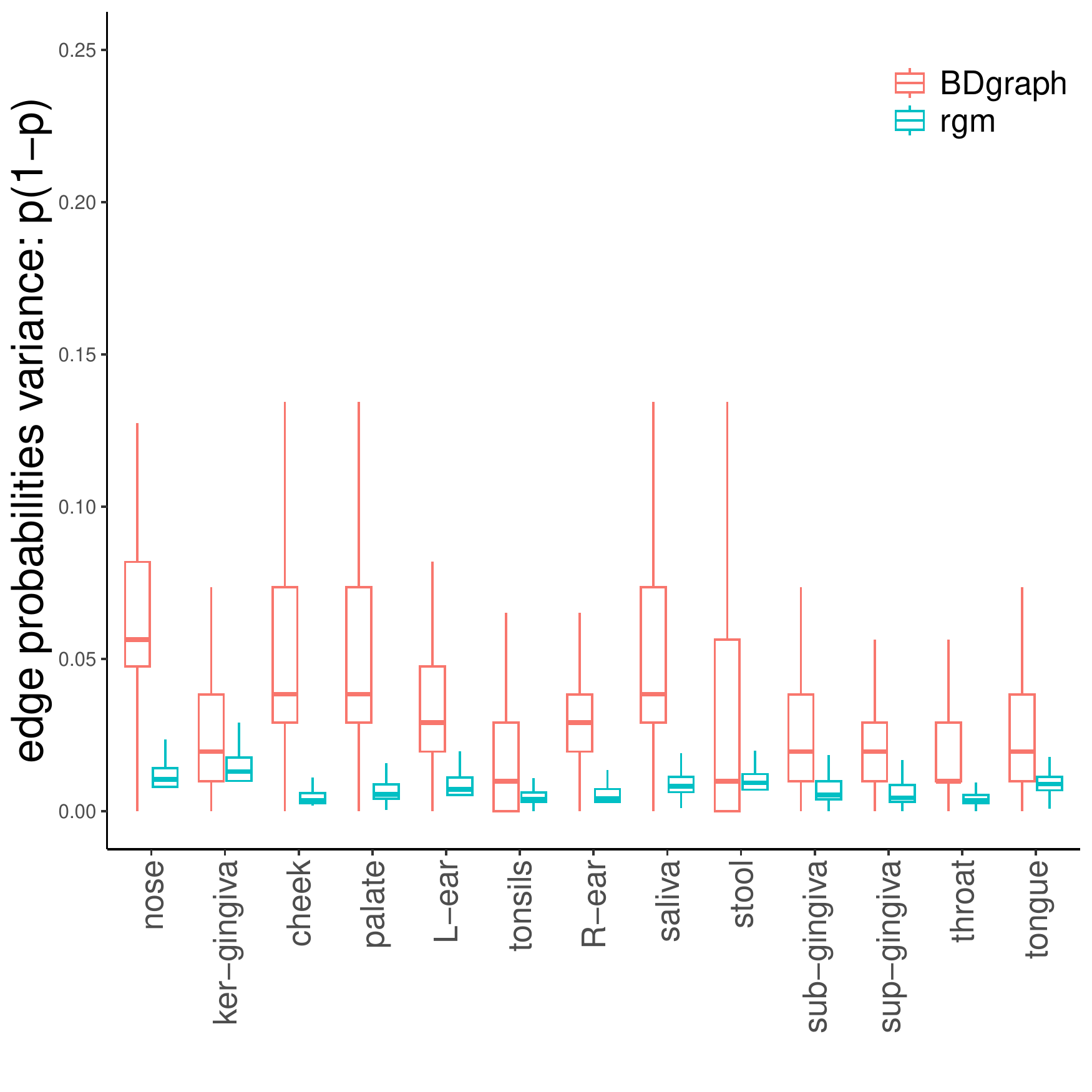}
\caption{Comparison between the joint \texttt{rgm} and individual Gaussian copula graphical models for each environment \citep{vinciotti22}. For each method and each environment, the plot shows the boxplot of the variance of the posterior edge probabilities.
\label{fig:06}}
\end{figure}

\vskip .1in
\noindent
{\bf Comparison with other methods.}  Figure~\ref{fig:06} shows that the random graphical model leads also to a more stable recovery of the individual networks, compared to estimating individual networks. Indeed, the figure shows that the variances of the posterior edge probabilities are smaller for the proposed \texttt{rgm} approach than when fitting individual Gaussian copula graphical models for each environment separately. For the latter, we use the approach of \cite{vinciotti22} (function \texttt{bdgraph.dw} in the \texttt{BDgraph} package), which uses the same parametric marginals as those considered in this paper but a more traditional Erd\"{o}s-R\'{e}nyi random graph prior for each environment. To facilitate comparison, the prior edge probability of the Erd\"{o}s-R\'{e}nyi prior is set to match the sparsity level of the networks recovered by the \texttt{rgm} analysis. Figure \ref{fig:06} shows how the posterior probabilities detected by \texttt{rgm} are more concentrated on either 0 or 1, than with the alternative approach, leading to a lower level of noise on the structural dependencies that are recovered by the method proposed in this paper.


\section{Discussion and Conclusion} \label{sec:conclusion}
In this paper, we have proposed a novel approach for the inference of microbiotic systems from multivariate measurements of microbial abundances across different, but related, environments. We have shown how the combination of graphical models for each environment with a joint random graph model describing the distribution of graphs across environments allows to learn about the individual microbiota systems as well as their structural similarities. In order to further adapt to the richness and complexity of microbiome data, the proposed approach allows for the inclusion of external covariates that may have an association with marginal microbial abundances or their interactions. 

Beyond the analysis presented in this paper, the method can be used more broadly on microbiome data measured across different conditions, where there is interest in learning structural dependencies within each environment and their similarities between environments. At this more general level, the proposed methodology share some similarities with graphical modelling approaches from data across multiple conditions, such as those described by \cite{ni22}. More dedicated random graph models may be needed depending on the context, e.g., for the case of microbiome data measured over time with dependencies that change over time.

\section*{Software}
R package available at https://github.com/franciscorichter/rgm

\section*{Funding}
EW and FRM acknowledge funding by the Swiss National Science Foundation (SNF 188334) 

\bibliographystyle{chicago}
\bibliography{bibfile}

\end{document}